\documentclass[10pt,journal,final,finalsubmission,twocolumn]{IEEEtran}
\usepackage{cite} 
\usepackage{graphicx}
\usepackage{epsfig,epstopdf}
\usepackage{subfigure}
\usepackage{url}
\usepackage{amssymb,amsmath,amsbsy,amsfonts}
\usepackage{expdlist}
\usepackage{color}
\usepackage{rotating}
\usepackage{multirow} 
\usepackage{tabularx}
\usepackage{pdflscape}
\usepackage{booktabs}
 \usepackage{bmpsize}
\usepackage{pgfplots}
 \usepackage{subfigure} 
\usepackage{pgf}
 \usepackage{tikz}
 \usepackage{algorithm}
\usepackage{algorithmic}
\usepackage{soul	}
\usepackage{float}
\floatstyle{plaintop}
\restylefloat{table}
\usepackage{bm}
\usetikzlibrary{arrows}
\DeclareGraphicsExtensions{.jpg,.pdf,.pdf,.gif,.pdf,.tex,.tikz}
\graphicspath{{./}}

\usetikzlibrary{plotmarks}
\usepackage{pgfplots}

\makeatletter
\def\addlegendimage{\pgfplots@addlegendimage}
\makeatother

\newcommand{\ve}[1]{\mathbf{#1}}

\def\miu{{\mathbf u}}
\def\miy{{\mathbf y}}
\def\miH{{\mathbf H}}

\def\miw{{\mathbf w}}

\usepackage{soul}
\def\np{\sigma_w^2}
\DeclareMathOperator{\diag}{diag}

\def\l{\boldsymbol{\lambda}}
\def\gam{\boldsymbol{\gamma}}

\def\I{\mathbf{I}}

\def\H{\mathbf{H}}

\def\u{\mathbf{u}}
\def\w{\mathbf{w}}
\def\y{\mathbf{y}}

\def\I{\mathbf{I}}

\def\g{\mathbf{g}}

\def\L{\boldsymbol{\Lambda}}

\long\def\symbolfootnote[#1]#2{\begingroup%
\def\thefootnote{\fnsymbol{footnote}}\footnote[#1]{#2}\endgroup}

\newcommand{\noteP}[1]{\textcolor{black}{#1}}

\IEEEoverridecommandlockouts
\begin{document}
%
\title{Probabilistic  MIMO Symbol Detection with Expectation Consistency Approximate Inference}
%
%


\author{\IEEEauthorblockN{Javier C\'espedes,\thanks{This work has been partly funded by the Spanish Government through projects MIMOTEX (TEC2014-61776-EXP), CIES (RTC-2015-4213-7), ELISA (TEC2014-59255-C3-3R)  and FLUID (TEC2016-78434-C3-3-R), by the Juan de la Cierva program (IJCI-2014-19150),  by the European Research Council (ERC) through the European Union’s Horizon 2020 research and innovation program under Grant 714161, and by Comunidad de Madrid (project 'CASI-CAM-CM', id. S2013/ICE-2845). We gratefully acknowledge the support of NVIDIA Corporation with the donation of the Titan X Pascal GPU used for this research. }}\thanks{J. C\'espedes, P. M. Olmos and M. S\'anchez-Fern\'andez are with the Signal Theory \& Communications Department, Universidad Carlos III de Madrid. Pablo M. Olmos is also with the Gregorio Mara\~n\'on Health Research Institute. E-mail: {\tt \{jcespedes,olmos,mati\}@tsc.uc3m.es}} \and
\IEEEauthorblockN{Pablo~M.~Olmos,} \and
\IEEEauthorblockN{Matilde~S\'{a}nchez-Fern\'{a}ndez}, \and \IEEEauthorblockN{Fernando Perez-Cruz\thanks{F. Perez-Cruz is with the Signal Theory \& Communications Department, Universidad Carlos III de Madrid. He is also the Chief Data Scientist at the Swiss Data Science Center (Switzerland). Email: {\tt fernando@tsc.uc3m.es}, {\tt fernando.perezcruz@sdsc.ethz.ch}}} 
}


\maketitle
\vspace{-0.5cm}

\begin{abstract}
In this paper we explore low-complexity  probabilistic algorithms for soft symbol detection in high-dimensional multiple-input multiple-output (MIMO) systems. We present a novel algorithm based on the Expectation Consistency (EC) framework, which describes the approximate inference problem as an optimization over a non-convex function. EC generalizes algorithms such as Belief Propagation and Expectation Propagation. For the MIMO symbol detection problem, we discuss feasible methods to find stationary points of the EC function and explore their tradeoffs between accuracy and speed of convergence. 
\noteP{ The accuracy is studied, first in terms of input-output mutual information and show that} the proposed EC MIMO detector greatly improves state-of-the-art methods, \noteP{with a complexity order cubic in the number of transmitting antennas}. \noteP{Second,} these gains are corroborated by combining the probabilistic output of the EC detector with a low-density parity-check (LDPC) channel code. 
\end{abstract}



\section{Introduction}

With the increasing demand for higher data rates, multiple-input multiple-output (MIMO) systems have attracted much attention over the last decade \cite{Mietzner09}. 
It is well known that MIMO communication systems achieve substantial gains in terms of spectral efficiency compared to conventional single-input single-output (SISO) systems. In fact, it has been shown that under ideal conditions the capacity of a point-to-point MIMO system with $m$ transmitting antennas and $r$ receiving antennas scales linearly with  $\min(m,r)$, which is referred to as the multiplexing gain \cite{Zheng03}.

Modern channel-coding techniques, such as Turbo codes \cite{bgt93} or  LDPC codes \cite{ru08}, are needed to achieve transmission rates close to the fundamental theoretical limits of the MIMO channel. Efficient decoding is possible using the belief propagation (BP) algorithm \cite{kfl01,ru08}, which is a low-complexity message-passing approximate inference method to estimate marginals in a joint probability distribution.  BP decoding needs as input an estimate to the  posterior probability of each coded bit given the vector of channel observations.  This information is provided by the so-called probabilistic symbol detector, which has to marginalize the posterior probability density function (pdf) of the transmitted vector of 
symbols, given the channel observation. For a MIMO channel, this has complexity $\mathcal{O}(M^m)$, where $M$ is the constellation order. 

%

Multiple algorithms have been proposed to perform hard-output symbol detection in MIMO systems, see \cite{Burg05,Guo06,Golden99,Liu08,Zhao07,Datta11,Zhou13,Caire-2004,Goldberger13,Hanzo15}. On the contrary, the  list of probabilistic  symbol detection algorithms is comparatively much shorter. Soft-ouput sphere decoding (SD) methods solve the marginalization in a sub-space of the constellation alphabet $\mathcal{A}^m$ \cite{Boutros03,Studer08,Wang06}. 
However, to maintain good performance,  the dimension of the sub-space must grow rapidly with $m$, the modulation order and the inverse of the signal-to-noise ratio (SNR)  \cite{Steingrimsson03}. \noteP{Thus, SD methods are not suitable for massive MIMO scenarios, where both $m$ and $M$ are potentially very large}.
Alternatively, some other works consider the use of Markov chain Monte Carlo (MCMC) algorithms to approximate the marginal posterior probabilities \cite{Datta13,Hansen09,Rong-Rong10,Jia04}. 
While this approach has been shown to be viable for hard-output symbol detection, probabilistic detection requires a sufficiently large number of samples \emph{per constellation point} at each transmitter. For large  $m$ and high-order constellations, MCMC methods become excessively burdensome.

\noteP{The focus of this paper is on} MIMO probabilistic symbol detection methods that can scale up to hundreds of antennas and high-order modulations based on quadrature amplitude modulation (QAM). In particular, we focus on methods with polynomial complexity with the number $m$ of transmit antennas. The minimum-mean-squared error (MMSE) solution can be cast as a probabilistic detector since it computes the mode of a Gaussian approximation to the posterior pdf of the MIMO symbols \cite{Caire-2004,Sanderovich05}, \noteP{likewise its soft successive interference cancellation (soft MMSE-SIC) version \cite{Wang07}. In both implementations} complexity is dominated by an $m\times m$ matrix inversion.  
The Gaussian tree approximation (GTA)  algorithm \cite{Goldberger11}, \noteP{very close in hard detection performance to MMSE-SIC,}  is a detection algorithm that constructs a tree-factorized approximation to posterior pdf of the symbols, to then estimate marginals distributions using BP. \noteP{Also, inspired by their success in compressed sensing \cite{Donoho13}, in recent years there has been an intense research interest on MIMO detection techniques based on message passing algorithms.} We can mention the 
channel hardening-exploiting message passing (CHEMP) in \cite{Narasimhan14} and the Gaussian Message Passing Iterative Detector  (GMPID) in \cite{Liu16ICC, Liu16}. Both methods have been shown to be effective (close to SD methods) for large MIMO systems with QPSK constellations. However,  asymptotic analysis of this type of algorithms shows that they do not perform well with high-order QAM constellations unless the number $r$ of receiving antennas is much larger than the number $m$ of transmitting antennas \cite{Jeon15,Liu16}. An improved version of the GMPID algorithm called SA-GMPID is shown to asymptotically converge to the  MMSE detection solution even for the case $m/r>1$ \cite{LiuGlobecom16}. We remark  that in this paper we propose algorithms that, while having larger complexity compared to these type of message-passing algorithms, they significantly improve the MMSE solution.

In \cite{Cespedes2014}, we proposed Expectation Propagation (EP) \cite{Minka01a,Seeger05} to perform hard-output MIMO symbol detection in the high SNR regime. In that paper, EP is used to find the mode of the posterior probability distribution by projecting it into a Gaussian approximation. The method cannot be easily \noteP{generalized} to perform probabilistic detection, as its description is essentially an iterative algorithm that does not provide the complete picture of the fundamental underlying inference problem. Actually, in \cite{Cespedes14ISIT} we showed that, while the MIMO EP receiver in  \cite{Cespedes2014} is able to significantly improve GTA as hard detector, achieving gains of around 2 dBs, both methods perform similarly  when combined with an LDPC channel decoder that requires a probabilistic input. In a simpler scenario, i.e. channel equalization for single-user intersymbol-interference (ISI) channels, different heuristics have been recently proposed in  \cite{Santos16}  to improve the EP probabilistic output, but it is shown that ultimately a turbo-like receiver, where the LDPC decoder output is fed back to the EP equalizer, is required to obtain a robust solution that is not tailored to a particular modulation or channel instance. 

In this work, \noteP{we consider one-shot receiver architectures, in which the channel decoder output is not fed back to the MIMO symbol detector to modify the original estimate. In this scenario, the design of the MIMO detector is particularly crucial, as the overall system performance highly depends on its accuracy. One-shot receivers can be used in latency-constrained applications instead of iterative Turbo-like receivers, as the latency in the latter case can become too large if long block channel codes are used \cite{Vitetta13}. } Furthermore, we show how probabilistic MIMO symbol detection can be implemented using a general approximate inference framework called Expectation Consistency (EC), which was first described by Opper \& Winther in \cite{Opper2005}. In EC, we describe the inference problem as the search of a stationary point of an approximation to the free energy associated to the true posterior probability distribution of the transmitted symbols. Any stationary point satisfies a  moment matching condition between the involved distributions. In this paper, we \noteP{tailor} the original EC formulation to the MIMO detection case and we discuss feasible methods to find such stationary points and show the fundamental tradeoffs between accuracy and speed of convergence. \noteP{In particular, we propose an update rule that performs very close to the moment matching  EC solution, with a complexity comparable to running MMSE ten times}.  \noteP{Also, we propose methods to overcome numerical instabilities that may arise in the MIMO detection scenario, particularly when we use large constellation alphabets.} In all tested scenarios, we find solutions that are robust and accurate across different modulation orders and system dimensions. Finally, the resulting EC probabilistic MIMO detector achieves excellent performance results compared to state-of-the-art methods with the same complexity order. 

To measure the accuracy of the \noteP{EC} MIMO detector probabilistic output, \noteP{first} we use a Monte Carlo estimate to the mutual information between the transmitted MIMO symbol vector and the corresponding output of the probabilistic symbol detection stage. At high SNRs, all detection methods saturate at the same mutual information level, i.e., $\log_2(M)$ bits per channel use per antenna, due to the use of a finite discrete constellation of $M$ points. Operating in the high-SNR region of saturation is  undesirable, as the gap to channel capacity grows exponentially as we increase the SNR. \noteP{However, at moderate SNR, our proposed detector outperforms other detectors in the literature and, in those scenarios where we could obtain the optimal detector solution, EC gets very close to it. Second,} the predicted gain at moderate-SNRs   is corroborated by bit error rate (BER) performance simulation using  optimized irregular LDPC block codes \cite{rsu01} and terminated convolutional-LDPC block codes \cite{Costello14,KudekarBMSIT}. In all cases, we obtain remarkable SNR gains, proving that the accuracy of the MIMO probabilistic symbol detection stage is crucial in the system's performance. 

Overall, the contributions of this paper are summarized as follows:
\begin{itemize}
\item \noteP{We introduce EC approximate inference framework and show how it can be applied to the MIMO detection scenario, developing the EC free energy approximation and computing its gradients.}
\item We compare several approaches to find EC stationary points, and propose iterative rules that are able to approach the optimal solution at $\mathcal{O}(m^3)$ complexity.
\item We \noteP{obtain} the achievable rate (mutual information) of a single-user MIMO system \noteP{to show the accuracy in the pdf approximation to  the true posterior, also proving} that with EC detection we significantly reduce the gap to capacity. The predicted gains are corroborated via error rate simulation with optimized LDPC codes.
\end{itemize}

The paper is structured as follows. In Section \ref{sec_channel_model} we  review the system model. In Section \ref{ratesection}, we discuss on the transmission rate and how it depends on the MIMO symbol detection method implemented, \noteP{highlighting the importance of a good approximation to the true posterior}. Section \ref{ECaproxInf} briefly presents the EC approximate inference framework and we \noteP{tailor} it to the MIMO detection case in Section \ref{algorithms}. In Section \ref{experimental_results},  experimental results are presented. Final conclusions and potential lines of  future research are described in Section \ref{conclusions}.

Notation: Capital and lowercase boldface symbols represent matrices and vectors respectively. $[\cdot]^\top$ is the transpose and $[\cdot]^{H}$ is the Hermitian. 
Finally, $[n]$ denotes the set $\{1,2,\ldots,n\}$.

\section{System Model}
\label{sec_channel_model}

\begin{figure}[t!]
\begin{center}
\includegraphics{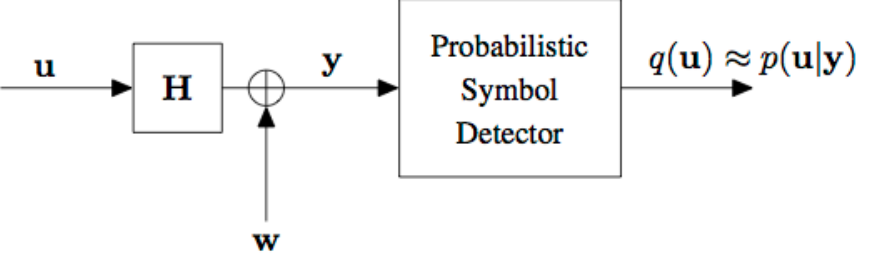}
\caption{System model}\label{modelo3}
\end{center}
\end{figure}

Consider a single-user MIMO system where $m$ transmitting antennas communicate to a receiver with $r$ antennas.  The system model is shown in Fig. \ref{modelo3}.   Let $\mathbf{b} = [b_{1}, b_2, ..., b_k]^\top$ denote the input information binary vector, which is Gray-mapped and modulated into  QAM symbols. Then, an $m$-dimensional vector of QAM symbols is generated, that is denoted by $\miu = \mathbf{u}_{\text{re}} + j\mathbf{u}_{\text{im}} \in \mathcal{A}^{m}$, where $|\mathcal{A}|=M$. The symbol vector $\miu$, is transmitted over a memoryless flat-fading complex MIMO channel, defined as a matrix $\miH$ with dimensions $r \times m$ of zero-mean unit-variance complex Gaussian coefficients.  Therefore, 
\begin{equation}
\miy = \miH\miu + \miw,
\label{channelmodel}
\end{equation}
where $\miy \in \mathbb{C}^r$ and $\miw \in \mathbb{C}^r$ is an additive white circular-symmetric complex Gaussian noise vector with independent zero-mean components and $\np$-variance. We  also assume  that the receiver has perfect channel state information (CSI). On the other hand, the signal-to-noise ratio is defined as
\begin{equation}\label{snr}
\text{SNR(dB)} = 10 \log_{10}\left(m \log_{2} (M) \frac{E_{b}}{\sigma_{w}^{2}}\right),
\end{equation}
where $E_b$ is the bit energy and the constellation energy $E_s$ can be written as 
\begin{equation}
E_s =E_b\log_2(M).
\end{equation}


\noteP{Note that the SNR defined is taking into account the full power transmission instead of the per-antenna power.}
Given the channel observation, the posterior distribution of the transmitted symbols, \noteP{that would lead to the optimal detector and that is also denoted through this work as true posterior, } is
\begin{align}\label{post}
p(\u|\y) = \frac{p(\y|\u)p(\u)}{p(\y)} \propto \mathcal{N}(\y:\H\u,\np\mathbf{I}) p(\u),
\end{align}
where \noteP{$\mathcal{N}(\y:\H\u,\np\mathbf{I})$ denotes a complex Gaussian} with mean $\H\u$ and covariance matrix $\np\mathbf{I}$, and $p(\u)$ is the prior probability density function for $\miu$. Assuming that we transmit independent uniformly distributed symbols, we have
\begin{equation}\label{prior}
p(\u) = \prod_{i=1}^{m}p(u_i) = \prod_{i=1}^{m} \frac{1}{M}\mathbb{I}_{u_i \in \mathcal{A}},
\end{equation}
where $\mathbb{I}_{u_i \in \mathcal{A}}$ takes value one if $u_i$ belongs to $\mathcal{A}$.  Observe that, due to the likelihood term in \eqref{post}, $p(\u|\y)$ is a multidimensional discrete distribution that maps over a fully
connected factor graph. Exact inference over $p(\u|\y)$, required to evaluate symbol marginals $p(u_i|\y)$, $i\in[m]$, to later feed a modern channel decoder,  has cost $\mathcal{O}(M^{m})$ and quickly (in both $M$ and $m$) becomes unfeasible. 

\subsection{Posterior approximation and inference}

\noteP{One of the alternatives to implement a low complexity} probabilistic symbol detector is to construct a tractable distribution $q(\u)$ that approximates $p(\u|\y)$.  By tractable we mean that performing inference over $q(\u)$, namely marginalizing it or computing expectations, is feasible. \noteP{Other options, reduce or modify the constellation space, as for example SD. }

\noteP{Focusing on the first alternative,} the MMSE method can be seen as a Gaussian approximation $q(\u)$ to $p(\u|\y)$ obtained by replacing the independent discrete priors in \eqref{prior} by the product of univariate zero-mean and $E_s$-variance complex circularly-symmetric Gaussian factors  \cite{Caire-2004,Sanderovich05}.  The Gaussian tree approximation (GTA) was first proposed in \cite{Goldberger11}. The method constructs a tractable cycle-free discrete approximation to \eqref{post} by replacing the Gaussian likelihood term $p(\y|\u)$ by a Gaussian distribution that factorizes in cycle-free graph, chosen to match the marginal and cross-moments of $p(\y|\u)$. Using this cycle-free approximation to the likelihood, efficient inference is carried out using BP.  Finally, there exist several recent proposals that perform approximate inference for MIMO symbol detection based on approximate message passing (AMP) \cite{Donoho13}. AMP algorithms essentially implement the standard rules of BP message passing \cite{Wainwright08} and  all messages are approximated with univariate Gaussian distributions. Among AMP methods for MIMO detection, we can mention the  CHEMP algorithm in \cite{Narasimhan14} and GMPID in \cite{Liu16}. An approximation to $p(\u|\y)$ can be constructed from the AMP marginals using the Bethe reparameterization \cite{Wainwright08}. 

\noteP{In Section \ref{complexity}, we have included a table summarizing the theoretical complexity order of each of the MIMO detection methods we use in our experiments.}

 

\section{Transmission rate}\label{ratesection}

Consider a fixed and known channel matrix $\H$\noteP{, under the system model defined in Section \ref{sec_channel_model}}. With the power constraint \noteP{$\mathbb{E}[\miu^T\miu]\leq \text{SNR}\sigma_{w}^{2}$}, 
the ergodic channel capacity per transmitted antenna with perfect CSI at the receiver and no CSI at the transmitter is given by
\begin{align}\label{eq:capacity}
C=\max_{p(\miu)} \frac{I(\miu,\miy)}{m}=\frac{\log_2(\det(\mathbf{I}_r+\frac{\text{SNR}}{m}\miH\miH^H))}{m}
\end{align}
bits per channel use and antenna. Capacity is achieved when $\miu$ is Gaussian distributed with zero-mean and covariance matrix equal to identity \cite{Telatar99}. 


\begin{figure}[t!]
\begin{center}
\includegraphics{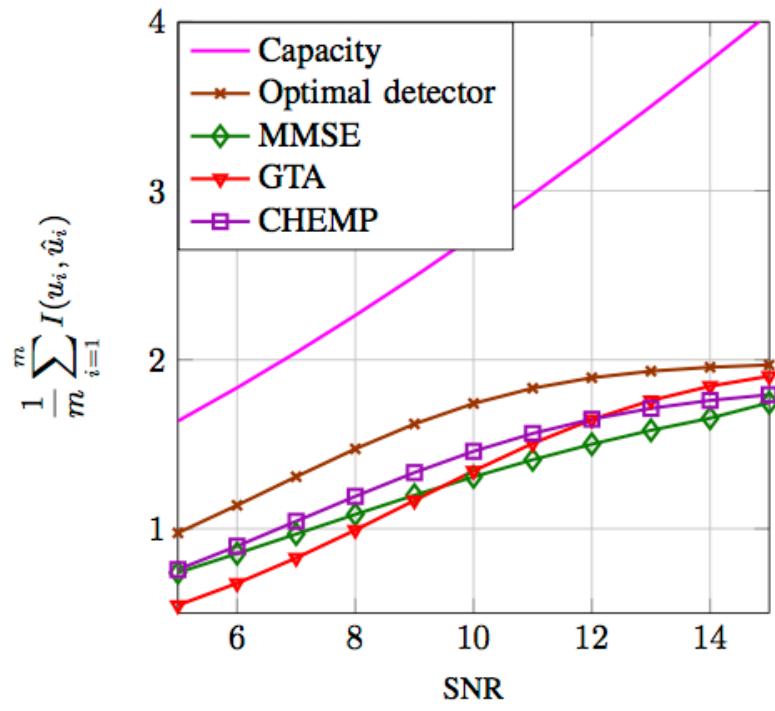}
\caption{Transmission rate  in $5\times 5$ scenario with QPSK modulation.}\label{inf}
\end{center}
\end{figure}

When $\miu$ is a random vector uniformly distributed in $\mathcal{A}^{m}$, the system transmission rate degrades and \noteP{can be far from the capacity limit} in \eqref{eq:capacity}. The achievable rate per antenna can be computed by evaluating the mutual information between $u_i$, the transmitted symbol at $i$-th antenna and $\hat{u}_i\sim p(u_i|\miy)$, i.e.,
\begin{equation}\label{eq:mutualinformation}
I(u_i,\hat{u}_i)=\mathbb{E}_{p(u_i,\hat{u}_i)}\left[\log_2\frac{p(\hat{u}_i|u_i)}{p(\hat{u}_i)}\right] \text{~~~(bits/channel use)},
\end{equation}
for  $i\in[m]$. Unfortunately, it is not possible to compute this mutual information in closed-form. We  follow a Monte Carlo procedure to estimate $\frac{1}{m}\sum_{i=1}^{m}I(u_i,\hat{u}_i)$ in the same channel knowledge scenario as the one assumed in \eqref{eq:capacity}, namely perfect CSI only at the receiver.  More precisely, we estimate $I(u_i,\hat{u}_i)$, $i\in[m]$, at one particular SNR point  as follows: first, we collect $N\in\mathbb{Z}_+$ samples from the joint distribution of $u_i,\miy$ and $\hat{u}_i$. Using this set of samples, we estimate $p(\hat{u}_i)$, $p(\hat{u}_i|u_i)$ for any \noteP{$u_i,\hat{u}_i\in\mathcal{A}$}, and, finally, compute a numerical estimate to $I(u_i,\hat{u}_i)$ in \eqref{eq:mutualinformation}.  As $N\rightarrow\infty$, the estimate to $I(u_i,\hat{u}_i)$ gets tight. Samples of the joint $(\miu,\miy,\hat{\miu})$ distribution are computed using \emph{ancestral sampling} \cite{Bishop06}. Each of the $N$ samples is generated following the next steps:
\begin{enumerate}
\item Sample $\miu$ from an uniform distribution in $\mathcal{A}^{m}$.
\item Sample $\miy$ from $p(\miy|\miu,\miH)$.
\item Sample $\hat{u}_i$, $i\in[m]$, from
\begin{align}\label{marg}
p(u_i|\miy)=\sum_{\u_{- i}} p(\miu|\miy) \qquad u_i \in \mathcal{A},
\end{align}
where $\u_{- i}$ denotes all elements in $\u$ except $u_i$.
\end{enumerate}
When a probabilistic symbol detector \noteP{does not use the true posterior}, the transmission rate can be evaluated by  following a similar procedure, but in 3) we sample  $\hat{u}_i$ after marginalization over $q(\miu)$, namely the approximation constructed to  $p(\miu|\miy)$. \noteP{Thus, the average mutual information computed for each low complexity detection method is used as a performance metric that measures how close $q(\miu)$ is to $p(\miu|\miy)$. At the same time,} the better the quality of the approximation is, the higher the rate becomes. \noteP{Note also that to compute this metric, we consider uncoded transmission.} For instance, Fig. \ref{inf} shows the average mutual information per antenna in a $5\times5$ scenario with QPSK modulation for both the optimal detector (which works directly with \noteP{the true posterior} $p(\miu|\miy)$), and for MMSE, GTA and CHEMP suboptimal detectors.  It has been computed with $N=10^6$ samples per SNR point. Also, results have been averaged over 100 realizations of $\H$. Observe that all methods operate close to the limit of $2$ bits/channel use when the SNR is high, but the gap to channel capacity in this regime grows exponentially fast with the SNR. For intermediate SNR values, optimal detection clearly outperforms MMSE, GTA and CHEMP detection\footnote{Note the similarities with the throughput results presented in \cite{Ketonen10}.} . It is precisely in this regime where we must improve the accuracy of the probabilistic symbol detection stage. 

\section{Expectation Consistency Approximate Inference for MIMO detection}\label{ECaproxInf}

In this section we give a brief introduction to EC approximate inference  \cite{Opper2005}, to then \noteP{tailor it for low-complexity} probabilistic MIMO detection. Let $\boldsymbol{U}$ be a random variable with a probability density function that factors in the following way
\begin{equation}\label{true2}
p(\u ) =\frac{1}{Z} f_{q}(\u )f_{r}(\u ),
\end{equation}
where we assume that computing  \noteP{$ Z = \int f_{q}(\u )f_{r}(\u )d\u$} or any expectation w.r.t. $p(\u)$ is unfeasible. However, we do assume that, separately, $f_{q}(\u )$ and $f_{r}(\u )$ are tractable w.r.t. a measure of the form $\exp(\l^T\boldsymbol{\phi}(\u))$ for some function vector $\boldsymbol{\phi}(\u)=[\phi_1(\u), \ldots,\phi_J(\u)]$. Namely, we assume it is possible to perform inference over the following two distributions used to approximate $p(\u )$:
\begin{align}\label{eq3}
q(\u) &= \frac{1}{Z_{q}(\l_q)}f_{q}(\u)\exp(\l_{q}^\top\boldsymbol{\phi}(\u )),\\ 
 r(\u ) &= \frac{1}{Z_{r}(\l_r)}f_{r}(\u)\exp(\l_{r}^\top\boldsymbol{\phi}(\u )),
\label{eq4}
\end{align}
where the \noteP{$J\times 1$} parameter vectors $\l_q$ and $\l_r$ belong to a certain convex set $\Phi$, and 
\begin{align}
Z_{q}(\l_q) &= \int f_{q}(\u )\exp(\l_{q}^\top\boldsymbol{\phi}(\u )) d\u \label{zq},\\
Z_{r}(\l_r) & = \int f_{r}(\u )\exp(\l_{r}^\top\boldsymbol{\phi}(\u )) d\u
\label{zr}.
\end{align}
Note that both $q(\u)$ and $r(\u)$ define an exponential family of distributions\footnote{See  \cite{Wainwright08}  for an introduction to exponential families and their properties.}, where $\l_q$ ($\l_r$) is the natural parameter vector, $\boldsymbol{\phi}(\u )$ is the vector of sufficient statistics and $\log Z_q(\l_q)$ ($\log Z_r(\l_r)$) is a convex function of $\l_q$ ($\l_r$) that satisfies
\begin{align}\label{grad1}
\nabla_{\l_q} \log Z_q(\l_q)=\mathbb{E}_{q(\u)}\left[\boldsymbol{\phi}(\u )\right],\\\label{grad2}
\nabla_{\l_r} \log Z_r(\l_r)=\mathbb{E}_{r(\u)}\left[\boldsymbol{\phi}(\u )\right].
\end{align}
The main idea behind EC approximate inference is to optimize $\l_q$ and $\l_r$ so that $q(\u)$ and $r(\u)$  have the same moments, i.e., \eqref{grad1} is consistent with \eqref{grad2}, keeping in mind that both $q(\u)$ and $r(\u)$, being the functions used to approximate $p(\u)$, contain ``partial information'' ($f_q(\u)$ and $f_r(\u)$ respectively) of this true distribution $p(\u)$. 

The first step to derive the EC approximation is to note that the partition function $Z$ in \eqref{true2} can be expressed the following way
\begin{align}
Z &= Z_{q}(\l_{q})\frac{Z}{Z_{q}(\l_{q})} = Z_{q}(\l_{q}) \frac{\displaystyle \int f_{q}(\u )f_{r}(\u )d\u }{Z_{q}(\l_{q})}=\nonumber \\
& = Z_{q}(\l_{q}) \displaystyle \int \frac{f_{q}(\u )}{Z_{q}(\l_{q})}f_{r}(\u )\exp((\l_{q} - \l_{q})^\top\boldsymbol{\phi}(\u ))d\u  \\
& = Z_{q}(\l_{q}) \mathbb{E}_{q(\u )}[f_{r}(\u )\exp(-\l_{q}^\top\boldsymbol{\phi}(\u ))].
\end{align}
And thus,
\begin{align}
\log Z = \log Z_{q}(\l_{q}) + \log\left(\mathbb{E}_{q(\u )}[f_{r}(\u )\exp(-\l_{q}^\top\boldsymbol{\phi}(\u ))]\right).
\label{lnzEC}
\end{align}
where $\log Z$ is also known as the \emph{energy function}. In order to estimate the expectation in the above expression, 
%
we replace $q(\miu)$ by a simpler distribution $s(\u)$ that belongs to the same exponential family than $q(\u)$ and $r(\u)$, i.e.,
\begin{equation}
s(\u ) = \frac{1}{Z_{s}(\l_{s})}\exp(\l_{s}^\top\boldsymbol{\phi}(\u )),
\label{aps}
\end{equation}
where $\log Z_{s}(\l_{s})$ is a convex function of $\l_s$ that satisfies $\nabla_{\l_s}\log Z_{s}(\l_s)=\mathbb{E}_{s(\u)}[\boldsymbol{\phi}(\u )]$.
While replacing $q(\u)$ by $s(\u)$ yields, in general, a poor approximation, it can be a fairly reasonable solution if both $q(\u)$ and $s(\u)$ have the same moments, namely if $\mathbb{E}_{q(\u)}\left[\boldsymbol{\phi}(\u )\right]=\mathbb{E}_{s(\u)}\left[\boldsymbol{\phi}(\u )\right]$. This condition is naturally achieved as a stationary point of the resulting approximation to $\log Z$. By replacing $q(\u)$ by $s(\u)$ in \eqref{lnzEC}, $\log Z$ is approximated by 
\begin{align}
& \log Z_{\text{EC}}(\l_q,\l_s)=\nonumber\\
&=\log Z_{q}(\l_{q}) + \log(\mathbb{E}_{s(\u)}[f_{r}(\u )\exp(-\l_{q}^\top\boldsymbol{\phi}(\u ))]),
\end{align}
and after simple manipulation this term can be  expressed as follows:
\begin{align}\label{ZEC}
&\log Z_{\text{EC}}(\l_q,\l_s)=\nonumber\\
&= \log Z_{q}(\l_{q}) + \log Z_{r}(\l_{s} - \l_{q})- \log Z_{s}(\l_{s}).
\end{align}
Recall that by assumption $Z_{q}(\l_{q})$, $Z_{r}(\l_{s} - \l_{q})$ and $Z_{s}(\l_{s})$ can be computed efficiently. And  note that $\log Z_{\text{EC}}$ depends only on $\l_q$ and $\l_s$, while it depends on three probability distributions: $q(\u)$ with parameter vector $\l_q$, $r(\u)$ with parameter vector $(\l_s-\l_q)$ and $s(\u)$ with parameter vector $\l_s$. Recall we seek moment matching between $q(\u)$ and $r(\u)$ and also between $q(\u)$ and $s(\u)$. While the first condition ensures that the two approximations that we construct to $p(\u)$ are consistent, the latter is required so that the measure replacement in the expectation in \eqref{lnzEC} is not too coarse. 
%
Both conditions are  satisfied at any point $(\l^*_q,\l^*_s)$ where the gradient of the EC energy function $\log Z_{\text{EC}}(\l_q,\l_s)$ is zero, i.e. optimization over $\log Z_{\text{EC}}(\l_q,\l_s)$ would lead to $(\l^*_q,\l^*_s)$.

\subsection{The EC free energy for MIMO detection}

To simplify the low-complexity detector derivation, we rewrite the probabilistic model in  \eqref{post} to work with real-valued distributions, considering the real $\mathcal{R}(\cdot)$ and imaginary $\mathcal{I}(\cdot)$ parts separately. Define $\widetilde{\mathbf{u}}=\begin{bmatrix}\mathbf{u}_{\text{re}}^\top & \mathbf{u}_{\text{im}}^\top\end{bmatrix}^\top$, $\widetilde{\mathbf{y}} = \begin{bmatrix}\mathcal{R}(\y)^\top & \mathcal{I}(\y)^\top\end{bmatrix}^\top$, $\widetilde{\mathbf{w}} = \begin{bmatrix}\mathcal{R}(\w)^\top & \mathcal{I}(\w)^\top\end{bmatrix}^\top$ and 
\begin{equation*}
\widetilde{\mathbf{H}}
=
\begin{bmatrix}
\mathcal{R}(\H) & -\mathcal{I}(\H) \\
\mathcal{I}(\H) & \mathcal{R}(\H) \\
\end{bmatrix}.
\end{equation*}
Thus, the real-valued channel model is
\begin{equation}
\widetilde{\mathbf{y}} = \widetilde{\mathbf{H}}\widetilde{\mathbf{u}} + \widetilde{\mathbf{w}},
\label{real}
\end{equation} 
where $\sigma_{\widetilde{w}}^{2} = \np/2$ is the variance of the real and imaginary parts of the noise and we define $\widetilde{\mathcal{A}}$ as the new alphabet for the real and imaginary components of the $M$-QAM constellation, $\widetilde{\u} \in \widetilde{\mathcal{A}}^{2m}$, with energy $\widetilde{E}_{s} = E_s/2$. In the rest of this work we adopt the real-valued channel model formulation in \eqref{real} and we drop the model indicator $\widetilde{(\cdot)}$ to keep the notation uncluttered. Therefore,  the a posteriori  probability pdf of the transmitted symbol vector $\u$, and that we propose to approximate with tractable pdfs, can be expressed as follows
\begin{align}
p(\u |\y )  = \frac{1}{Z} \mathcal{N}(\y :\H \u,\sigma_w^{2}\I)\prod_{i=1}^{2m} \mathbb{I}_{u_i \in \mathcal{A}},
\label{eqbase}
\end{align}

The matching of \eqref{eqbase} with functions $f_q(\u)$ and $f_r(\u)$ in \eqref{true2} will be done so that  $q(\u)$ and $r(\u)$ in \eqref{eq3} and \eqref{eq4} are tractable w.r.t. a measure of the form $\exp(\l^T\boldsymbol{\phi}(\u))$, which means that we have to be able to easily compute moments of the form $\mathbb{E}[\boldsymbol{\phi}(\u)]$ w.r.t. both distributions. For an EC based low-complexity detector we choose the vector of statistics and natural parameters as follows
\begin{align}\label{moments}
&\boldsymbol{\phi}(\u ) = \left[u_1, u_2, \ldots, u_{2m},\frac{-u_1^2}{2}, \frac{-u_2^2}{2}, \ldots, \frac{-u_{2m}^2}{2}\right]^\top,\\ \label{params}
&\l =\left[\gamma_1,\gamma_{2},\ldots,\gamma_{2m}, \Lambda_1,\Lambda_2,\ldots,\Lambda_{2m}\right]^\top=[\gam,\L]^\top,
\end{align} 
where $\gam\in\mathbb{R}^{2m}$ and $\L\in\mathbb{R}^{2m}_{+}$. According to \eqref{moments}, this choice of $\boldsymbol{\phi}(\u)$ implies that at any zero-gradient point of the EC energy function in \eqref{ZEC}, the distributions 
$q(\u)$ and $r(\u)$ must be consistent only in their marginal first and second order moments. Under this assumption, if we choose functions $f_q(\u)$ and $f_r(\u)$ as follows
\begin{align}\label{funcs}
f_q(\u)=\mathcal{N}(\y :\H \u,\sigma_w^{2}\I),  \text{ and}~ f_r(\u)=\prod_{i=1}^{2m} p(u_{i})
\end{align}
then we conclude that $q(\u)$ and $r(\u)$ are tractable probability density functions, since $q(\u)$ is a Multivariate Normal distribution and $r(\u)$ is a discrete independent distribution. More precisely, according to \eqref{eq3} and \eqref{funcs}, we have
\begin{align}
&q(\u )=\frac{1}{Z_q(\gam_q,\L_q)} f_{q}(\u )\exp\left(\gam_{q}^\top\u  - \frac{\u ^\top\diag(\L_{q})\u }{2}\right) \nonumber\\\label{qMIMO}
&=\frac{\exp \left( \underbrace{\left( \frac{\H^\top\y }{\sigma_w^{2}} + \gam_{q}\right)^\top}_{\g^\top}\u - \frac{1}{2}\u ^\top\underbrace{\left(\frac{\H^\top\H}{\sigma_w^{2}} + \diag(\L_{q}) \right)}_{\mathbf{S}}\u\right)}{Z_q(\gam_q,\L_q)},
\end{align}
where $\diag(\L_q)$ is a diagonal matrix with main diagonal given by $\L_q$. Therefore $q(\u)=\mathcal{N}(\u: \bm{\mu},\bm{\Sigma})$, and $\bm{\Sigma} = \mathbf{S}^{-1}$ and $\bm{\mu} = \mathbf{S}^{-1}\g$. Also, we obtain
\begin{align}\label{logzq}
\log Z_q(\gam_q,\L_q) &= \frac{1}{2}\bm{\mu}^T\bm{\Sigma}^{-1}\bm{\mu}^T+\frac{1}{2}\log |\bm{\Sigma}|.
\end{align}
By applying standard rules for matrix derivatives, we can check that 
\begin{align}
\frac{\partial \log Z_q(\gam_q,\L_q)}{\partial \gamma_{q,i}} &= \mathbb{E}_q[u_i]= \mu_i,\\
\frac{\partial \log Z_q(\gam_q,\L_q)}{\partial  \Lambda_{q,i}} &= -\frac{1}{2} \mathbb{E}_q[u^2_i]= -\frac{1}{2}\left(\Sigma_{ii}+\mu^2_i\right).
\end{align}
On the other hand, from the definition of $f_r(\u)$ in \eqref{funcs} we get
\begin{align}
r(\u)&=\frac{1}{Z_r(\gam_r,\L_r)}  \exp\left(\gam^T_r\u-\frac{\u^T\diag\left({\L_r}\right)\u}{2}\right)\prod_{i=1}^{2m} \mathbb{I}_{u_i \in \mathcal{A}}\nonumber\\
&= \frac{1}{Z_r(\gam_r,\L_r)}  \prod_{i=1}^{2m} \exp\left(\gamma_{ri}u_{i}-\frac{\Lambda_{ri} u_{i}^2}{2}\right) \mathbb{I}_{u_i \in \mathcal{A}}\label{rMIMO}.
\end{align}
Therefore, $r(\u)$ is an independent discrete pmf over $\mathcal{A}^{2m}$ such that, for $i\in [2m]$,
\begin{align}\label{mur}
\mathbb{E}_r[u_i]&=\frac{\sum_{u_i\in \mathcal{A}} u_i \exp\left(\gamma_{ri}u_{i}-\frac{\Lambda_{ri} u_{i}^2}{2}\right)}{{\sum_{a\in \mathcal{A}} \exp\left(\gamma_{ri}a-\frac{\Lambda_{ri} a^2}{2}\right)}},\\\label{mur2}
\mathbb{E}_r[u^2_i]&=\frac{\sum_{u_i\in \mathcal{A}} u^2_i \exp\left(\gamma_{ri}u_{i}-\frac{\Lambda_{ri} u_{i}^2}{2}\right)}{{\sum_{a\in \mathcal{A}} \exp\left(\gamma_{ri}a-\frac{\Lambda_{ri} a^2}{2}\right)}}.
\end{align}
Also we have
\begin{align}
\log Z_r(\gam_r,\L_r) = \log \left(\sum_{a\in \mathcal{A}} \exp\left(\gamma_{ri}a-\frac{\Lambda_{ri} a^2}{2}\right)\right),
\end{align}
where we can again check that, $\frac{\partial \log Z_r(\gam_r,\L_r)}{\partial \gamma_{r,i}} = \mathbb{E}_r[u_i]$ and $\frac{\partial \log Z_r(\gam_r,\L_r)}{\partial \Lambda_{r,i}} = -\frac{1}{2}\mathbb{E}_r[u^2_i]$, for $i\in[2m]$. Finally, the averaging distribution $s(\u)$ in \eqref{aps} is given by
\begin{align}\label{sMIMO}
s(\u) = \frac{1}{Z_{s}(\l_{s})} \exp\left(\gam^\top_s\u-\frac{\u^\top\diag\left(\L_s\right)\u}{2}\right),
\end{align}
and therefore $s(\u)$ is an independent Gaussian distribution, i.e. $s(\u) =\mathcal{N}(\u: \diag\left(\L_s^{-1}\right)\gam_s, \diag\left(\L_s^{-1}\right))$. 

\noteP{Note that, given the vector of moments in \eqref{moments}, any choice for the functions $f_q(\u)$ and $f_r(\u)$ different to \eqref{funcs}, where some discrete priors are multiplied together with the Gaussian likelihood term $p(\y|\u)$, would result in $q(\u)$ or $r(\u)$ being an hybrid distribution, with some components taking values only in $\mathcal{A}$ and some other components taking real values. In such a case, evaluating the moments $\mathbb{E}[\boldsymbol{\phi}(\u)]$ would be an issue. On the other hand, while many other statistics can be included in the vector $\boldsymbol{\phi}(\u)$, e.g.  cross moments of the form $u_iu_j$ for some or all pairs of variables, we will show in the experimental results session that our choice in \eqref{moments} drives a robust and accurate MIMO detector. For instance, in the experimental section we show that the EC-based MIMO detector average mutual information in \eqref{eq:mutualinformation} is very close to the optimal detector for an scenario where the true posterior can be evaluated. Hence, there is little room for improvement of the EC solution by including higher order moments in $\boldsymbol{\phi}(\u)$. }


\section{Optimizing the MIMO EC free energy}\label{algorithms}

As described in the previous section, the goal in EC inference is to find $(\gam_q,\L_q)$ and $(\gam_s,\L_s)$ such that $q(\u)$ in \eqref{qMIMO}, $r(\u)$ in \eqref{rMIMO} (evaluated at $\gam_r=\gam_s-\gam_q$ and $\L_r=\L_s-\L_q$) and $s(\u)$ in \eqref{sMIMO} satisfy
\begin{align}\label{mm1}
\mathbb{E}_q[u_i]&=\mathbb{E}_r[u_i]=\mathbb{E}_s[u_i]\\\label{mm2}
\mathbb{E}_q[u^2_i]&=\mathbb{E}_r[u^2_i]=\mathbb{E}_s[u^2_i]
\end{align}
for $i\in[2m]$.  

\noteP{To achieve such a point, we present two algorithms. The so-called \emph{single loop} (SL), iteratively updates either $(\gam_q,\L_q)$ or $(\gam_r,\L_r)$ and follows a message-passing procedure. The resulting algorithm has approximately the MMSE complexity per iteration (see Table \ref{complexity_table}). On the other hand, by exploiting the fact that the EC free energy in $\eqref{ZEC}$ is a convex function w.r.t. $(\gam_q,\L_q)$, the so-called \emph{double loop} algorithm (DL) performs iteratively a convex optimization to set $(\gam_q,\L_q)$ for fixed $(\gam_s,\L_s)$ to then update the latter. Simulation results in Section \ref{convergence} show that the DL algorithm typically converges to a point closer to the stationarity conditions in \eqref{mm1}-\eqref{mm2}. As a caveat, its complexity is extremely large (see Table \ref{complexity_table}) and we would rather use it as a benchmark to improve the single loop approach.}

\noteP{It is important to remark that, for both algorithms, convergence to   \eqref{mm1}-\eqref{mm2} is not guaranteed \cite{Opper2005}. Actually, in most cases we observe that both algorithms get stuck in a  $(\l_q,\l_r)$ point for which these parameters do not change anymore  but at the same time the moment matching (MM) condition is not fully met. Our goal is to design robust algorithms to optimize the EC free energy such that they converge to stable $(\l_q,\l_r)$ points that are as close to the MM condition as possible.}

\begin{algorithm}[b]
\begin{algorithmic}
\STATE
Fix a damping factor $\beta$. Set maximum number of iterations $I_{\text{EC-S}}$. Set $\ell=0$.
\STATE
Initialize $\gam_q^{(0)}=\ve{0}$ and $\Lambda_{qi}^{(0)}=E_s^{-1}$ $i\in[2m]$. 
\REPEAT 
\STATE
1) Given $\gam_q^{(\ell-1)},\L_q^{(\ell-1)}$, compute $\mathbb{E}_{q}[u_i]$ and $\mathbb{E}_{q}[u_i^2]$, $i\in[2m]$.
\STATE
2) Compute $\gam_s^{(\ell)},\L_s^{(\ell)}$ such that $\mathbb{E}_s[u_i] = \mathbb{E}_{q}[u_i]$ and $\mathbb{E}_s[u^2_i] = \mathbb{E}_{q}[u^2_i]$, $i\in[2m]$.
\STATE
3) Update $\gam_r^{(\ell)}=\gam_s^{(\ell)}-\gam_q^{(\ell)}$, $\L_r^{(\ell)}=\L_s^{(\ell)}-\L_q^{(\ell)}$.
\STATE
4) Given $\gam_r^{(\ell)},\L_r^{(\ell)}$, compute $\mathbb{E}_{r}[u_i]$ and $\mathbb{E}_{r}[u_i^2]$, $i\in[2m]$.
\STATE
5) Compute $\gam_s^{(\ell)},\L_s^{(\ell)}$ such that $\mathbb{E}_s[u_i] = \mathbb{E}_{r}[u_i]$ and $\mathbb{E}_s[u^2_i] = \mathbb{E}_{r}[u^2_i]$, $i\in[2m]$.
\STATE
6)  Update 
\begin{align*}
\gam_q^{(\ell)}=\beta\left(\gam_s^{(\ell)}-\gam_r^{(\ell)}\right)+(1-\beta)\gam_q^{(\ell-1)}\\
\L_q^{(\ell)}=\beta\left(\L_s^{(\ell)}-\L_r^{(\ell)}\right)+(1-\beta)\L_q^{(\ell-1)}
\end{align*}
\STATE
7)  $\ell=\ell+1$
\UNTIL{convergence (or $\ell>I_{\text{EC-S}}$)}
\end{algorithmic}
\caption{The EC MIMO detector with SL updates}\label{SingleLoopMIMO}
\end{algorithm}

\subsection{The EC MIMO detector with single loop updates}

We initialize $(\gam_q,\L_q)$ such that $q(\u)$ in \eqref{qMIMO} coincides with the MMSE Gaussian approximation, i.e., $\gam_q^{(0)}=\ve{0}$ and $\Lambda_{qi}^{(0)}=E_s^{-1}$  $\forall i\in[2m]$\cite{Caire-2004,Sanderovich05}. The main steps are summarized Algorithm \ref{SingleLoopMIMO}. The complexity per  iteration is dominated by the computation of the covariance matrix of the $q(\u)$ distribution in \eqref{qMIMO} at step 1) of the algorithm. This complexity is $\mathcal{O}(m^3)$, but independent on the constellation size $M$. After the matrix inversion, computing the mean of $q(\u)$ requires $\mathcal{O}(m^2)$ operations. Computing the $r(\u)$ mean and variance in \eqref{mur} and \eqref{mur2} requires $\mathcal{O}(mM)$ operations. The complexity of the rest of steps does not depend on the constellation and thus the complexity is $\mathcal{O}(m)$.  Therefore, if  the  algorithm is run for $I_{\text{EC-S}}$ iterations, the final complexity is $\mathcal{O}(m^3I_{\text{EC-S}}+m^2 I_{\text{EC-S}}+mMI_{\text{EC-S}}+mI_{\text{EC-S}})$. 

\noteP{Numerical issues arise  due to the fact that we are propagating moments between a continuous and a discrete distribution, particularly in scenarios where all the mass of the marginal $r(u_i)$ distribution is concentrated in a small region of a potentially very large QAM constellation. This leads to small values of the marginal variance {\color{black}$\text{Var}_{r}[u_i]$} and, consequently,  $\Lambda_{si}$ may diverge in step 5). In order to avoid numerical issues, we implement a \emph{damping} (low-pass filter) in the update of $(\gam_q,\L_q)$ at step 6) of Algorithm \ref{SingleLoopMIMO}. Smoothing parameter updates via damping is a fairly common technique to stabilize approximate inference iterative algorithms. See for instance \cite{Heskes03,MooijKappen_IEEETIT_07,Elidan11} for discussions on message-passing stabilization.  }

\begin{algorithm}[b]
\begin{algorithmic}
\STATE
Fix a damping factor $\beta$. Set maximum number of iterations $I_{\text{EC-D}}$. Set $\ell=0$.
\STATE
Initialize $\gam_s^{(0)}=\ve{0}$ and $\Lambda_{si}^{(0)}=E_s^{-1}$ $i\in[2m]$. 
\REPEAT 
\STATE
1) Given $\gam_s^{(\ell-1)},\L_s^{(\ell-1)}$, solve the convex optimization in \eqref{convex}.
\STATE
2) Compute $\gam_s^{(\ell)},\L_s^{(\ell)}$ such that $\mathbb{E}_s[u_i] = \mathbb{E}_{q}[u_i]$ and $\mathbb{E}_s[u^2_i] = \mathbb{E}_{q}[u^2_i]$, $i\in[2m]$.
\STATE
3)  Update 
\begin{align*}
\gam_s^{(\ell)}=\beta\left(\gam_s^{(\ell)}\right)+(1-\beta)\gam_s^{(\ell-1)}\\
\L_s^{(\ell)}=\beta\left(\L_s^{(\ell)}\right)+(1-\beta)\L_s^{(\ell-1)}
\end{align*}
\STATE
4)  $\ell=\ell+1$
\UNTIL{convergence (or $\ell>I_{\text{EC-D}}$)}
\end{algorithmic}
\caption{The EC MIMO detector with DL updates}\label{DoubleLoopMIMO}
\end{algorithm}

\subsection{The EC MIMO detector with double loop updates}
\noteP{
The double loop algorithm is based on a simultaneous update of both $q(\u)$ and $r(\u)$ at every iteration by solving the following convex optimization problem for a fixed $(\gam_s,\L_s)$
\begin{align} \label{convex}
&(\gam_q^*,\L_q^*)=\arg\min_{(\gam_q,\L_q)} \log Z_{\text{EC}}(\gam_q,\gam_s,\L_q,\L_s)\\
&=\arg\min_{(\gam_q,\L_q)} \left(\log Z_q(\gam_q,\L_q)+ \log Z_r(\gam_s-\gam_q,\L_s-\L_q)\right)\nonumber
\end{align}
At $(\gam_q^*,\L_q^*)$, both $q(\u)$ and $r(\u)$ have the same moments. Then, $(\gam_s,\L_s)$ is recomputed to enforce moment matching (as in step 2) of Algorithm \ref{SingleLoopMIMO}). Instead of using the distribution $s(\u)$ to iteratively communicate the moments between $q(\u)$ and $r(\u)$, as the single loop algorithm does, note that the double loop is directly optimizing together both $q(\u)$ and $r(\u)$ to then update $s(\u)$. The main steps are outlined in Algorithm \ref{DoubleLoopMIMO}. We could use standard gradient descend to numerically solve \eqref{convex} in step 1). Note that in \eqref{logzq}, evaluating the gradient of $\log Z_q(\gam_q,\L_q)$ w.r.t. $(\gam_q, \L_q)$, requires a matrix inversion and a matrix product and thus a complexity of $\mathcal{O}(m^3+m^2)$. If $D$ denotes the number of gradient descend steps and $I_{\text{EC-D}}$ is the number of iterations, then the complexity is $\mathcal{O}(m^3DI_{\text{EC-D}}+m^2DI_{\text{EC-D}} + mI_{\text{EC-D}})$. }

\subsection{Assessing convergence}\label{convergence}

The moment matching condition in \eqref{mm1} and \eqref{mm2} represents the optimal operational point of the EC approximation. We emphasize that this notion of optimality is measured in terms of moment matching between tractable approximations to $p(\u|\y)$ ($q(\u)$ and $r(\u)$ respectively), and not w.r.t. the distribution $p(\u|\y)$ itself. 
%
%

For our experiments, we \noteP{study} the evolution of the following two quantities along iterations of the single loop EC MIMO detector:
\begin{align}
\Delta_u&={\color{black}\frac{1}{2m}}\sum_{i=1}^{2m} \Big| \mathbb{E}_q[u_i]-\mathbb{E}_r[u_i]\Big|,\\
\Delta_{u^2}&={\color{black}\frac{1}{2m}}\sum_{i=1}^{2m} \Big| \mathbb{E}_q[u^2_i]-\mathbb{E}_r[u^2_i]\Big|.
\end{align}

\begin{figure}[bh]
\begin{center}
\begin{tabular}{c}
\includegraphics{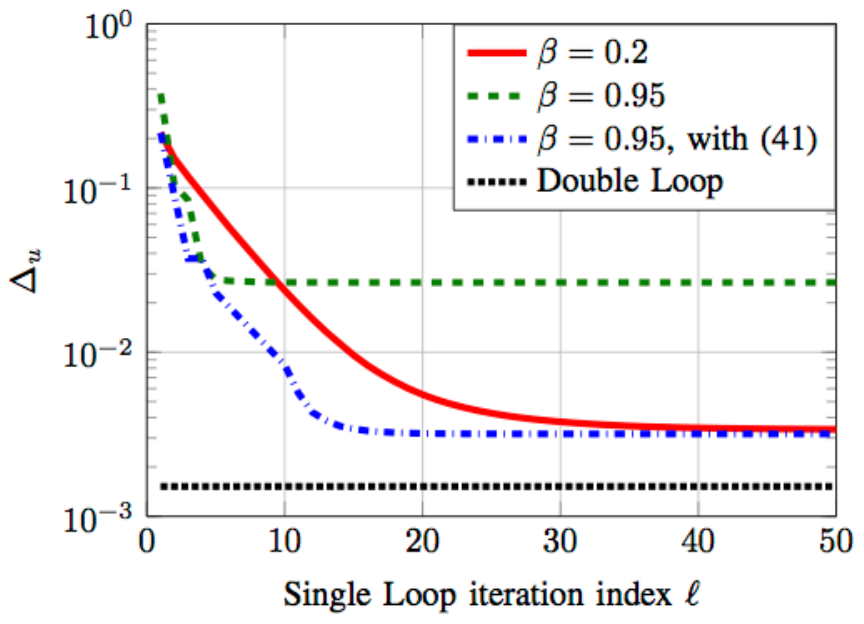} \\ (a) \\ \includegraphics{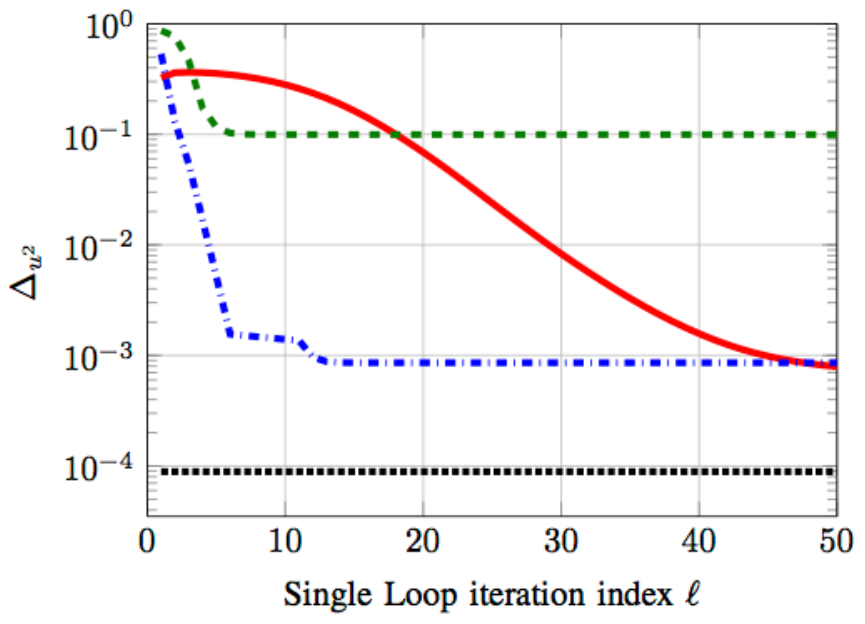}\\
(b)
\end{tabular}
\end{center}
\caption{We represent $\Delta_u$ and $\Delta_{u^2}$ for an $5\times 5$ scenario with QPSK modulation at a SNR of $6$dB, averaged over $10^4$ realizations of both the channel matrix $\H$ and received vector $\y$.}\label{conv1}
\end{figure}

\begin{figure*}[th]
\begin{center}
\begin{tabular}{cc}
\includegraphics{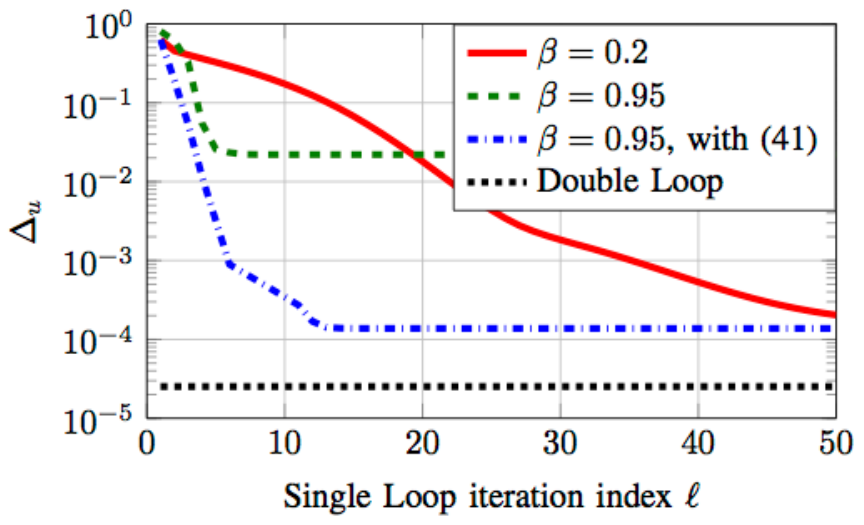} & \includegraphics{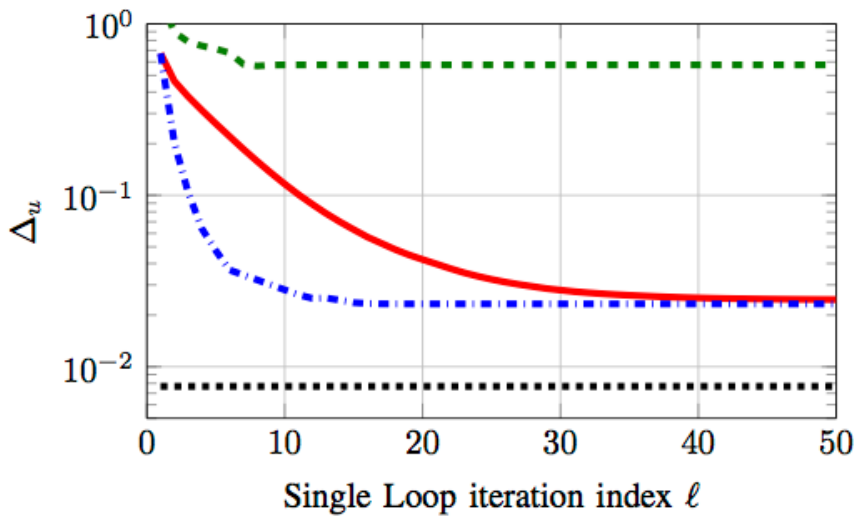} \\
(a) & (c) \\
\includegraphics{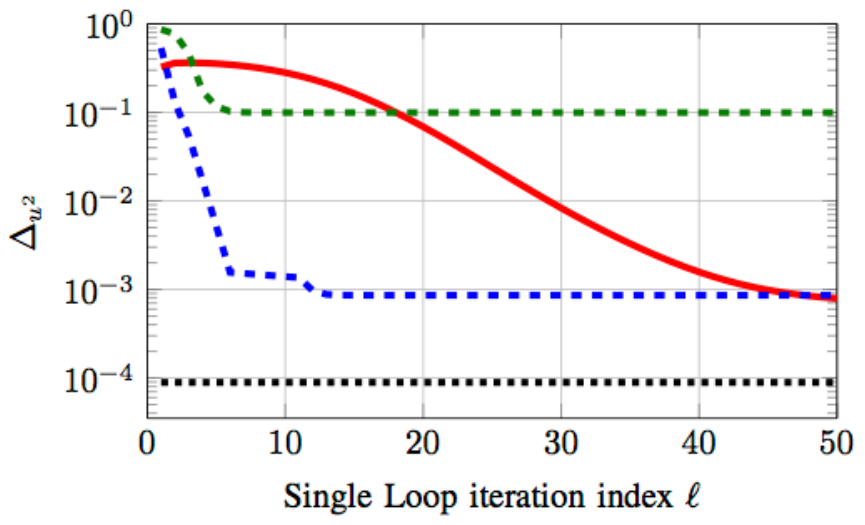} & \includegraphics{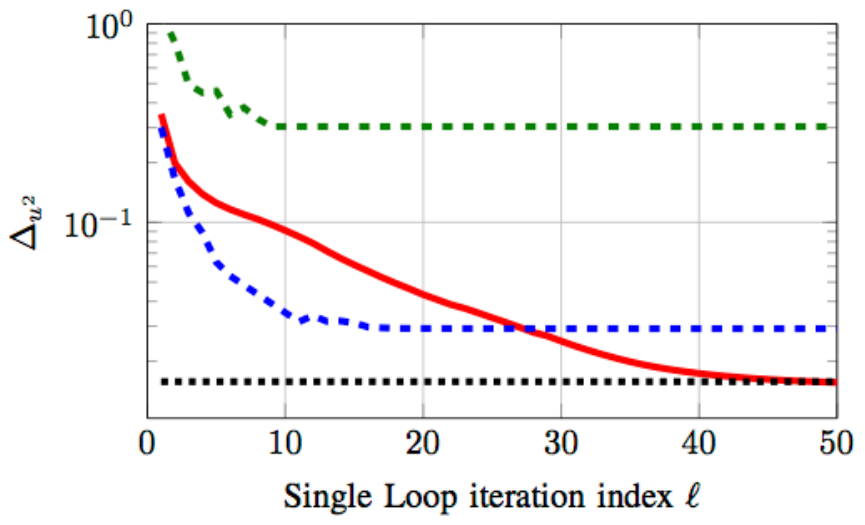} \\
(b) & (d)
\end{tabular}
\caption{In (a)-(b), we represent $\Delta_u$ and $\Delta_{u^2}$ for an $32\times 32$ scenario with QPSK modulation  at a SNR of $6$dB, averaged over $10^4$ realizations of both the channel matrix $\H$ and received vector $\y$. In (c)-(d), we reproduce the results for an $32\times 32$ scenario with 64-QAM modulation at a SNR of $21$dB.}\label{32conv}
\end{center}
\end{figure*}

In Fig. \ref{conv1} we represent $\Delta_u$ and $\Delta_{u^2}$ for a $5\times 5$ scenario with QPSK modulation at a SNR of $6$dB, averaged over $10^4$ realizations of both the channel matrix $\H$ and received vector $\y$.  According to Fig. \ref{inf}, this SNR value is far from the saturation regime (largest gap to channel capacity), and it is in this range where we aim the EC detector at substantially improving state-of-the-art methods. \noteP{With dotted black line we represent the double loop benchmark, computed for $I_{\text{EC-D}}=50$ iterations. At every iteration, we found that $D$, the number of gradient descend updates at step 1) of  Algorithm \ref{DoubleLoopMIMO}, has to be to a very large value until the gradient norm was below a  threshold of $0.1$. We set an upper limit of $D=2000$ and a gradient descend step-size of $10^{-3}$. We remark that every gradient descend step is as complex as a single iteration of the single loop EC algorithm.}

Three implementations of the SL algorithm are compared in Fig. \ref{conv1}. For the red solid line we have used $\beta=0.2$, i.e., a very slow parameter update in step 6) of Algorithm \ref{SingleLoopMIMO}. The opposite case is represented by the green dashed line, which has been computed with $\beta=0.95$. While the $\beta=0.2$ case approaches the double loop solution, achieving $\Delta_u$ and $\Delta_{u^2}$ around $10^{-3}$, it requires in average 25 iterations to converge to such a stationary point. Recall that each single loop iteration is as complex as computing the MMSE estimate, due to the matrix inversion in \eqref{qMIMO}. On the other hand, the $\beta=0.95$ case quickly saturates (around 10 iterations), but its solution is still far from the MM condition.

In order to achieve a better trade-off between accuracy and complexity, we maintain the fast updates using $\beta=0.95$, but modify the parameter update in Algorithm \ref{SingleLoopMIMO} and introduce a gradual decrease in the variance per component allowed at each iteration. More precisely, we set an iteration-dependent minimum value of the variance $\mathbb{E}_s[u^2_i]$ at step 5) of Algorithm \ref{SingleLoopMIMO} of the following form:
\begin{align}\label{minvar}
\text{Var}_s[u_i]=\max\left(2^{-\max(\ell-4,1)},\text{Var}_r[u_i]\right),
\end{align}
namely during the first $5$ iterations we set a reasonably minimum high variance per component ($0.5$) and, from iteration $4$, we let this minimum value to decrease exponentially fast with $\ell$. The convergence of this implementation of the EC algorithm is represented in Fig. \ref{conv1} with blue dashed-dotted lines. Observe that an improvement is achieved w.r.t. the $\beta=0.95$ case, reducing the gap w.r.t. to the stationary point achieved by $\beta=0.2$, without a significant penalty in speed of convergence, as it typically converges in less than 10 iterations. These effects are even more evident when we move to higher-dimensional scenarios. In Fig. \ref{32conv}  we consider a  $32\times 32$ scenario with QPSK (a)-(b) and 64-QAM modulation (c)-(d). Convergence speed is actually maintained and the gap w.r.t. the $\beta=0.2$ case is clearly reduced.  \noteP{While the parameter update in \eqref{minvar} was obtained heuristically after an intense empirical evaluation of the algorithms, we interpret the improvement achieved as follows. Setting a high-variance parameter during the first iterations of the algorithm is crucial in the low-SNR regime in order to avoid over-fitting. For large values of $\beta$, we observed that the single loop EC algorithm performance is degraded by very small values of the $r(u_i)$ variance ($\text{Var}_{r}[u_i]$) at  early iterations (step 4) of Algorithm \ref{SingleLoopMIMO}, indicating a very peaky distribution around a small region of the QAM constellation. Note that a very small variance is propagated to the $s(u_i)$ distribution at step 5) of Algorithm \ref{SingleLoopMIMO} with very large values of $\Lambda_{si}$. According to \eqref{sMIMO}, we have
\begin{align}
\Lambda_{si}^{-1}=\text{Var}_{s}[u_i]=\mathbb{E}_{r}[u^2_i]-\left(\mathbb{E}_{r}[u_i]\right)^2=\text{Var}_{r}[u_i],
\end{align}
and the same effect is propagated to $\Lambda_{qi}$ at step 6) of the algorithm unless $\beta$ is small enough. Very large values of $\Lambda_{qi}$ will dominate the diagonal of the matrix in \eqref{qMIMO} and, ultimately, this implies that successive steps of the EC algorithm will not be able to significantly change the $u_i$ marginal distribution anymore. Note that this is dramatic to the algorithm performance if the mode of the $r(u_i)$ distribution is placed at the wrong symbol, which is likely to happen at high-noise levels.}

\begin{table}[t!]
\small
\centering
\begin{tabular}{l|c}
\hline
MIMO detector & Complexity order\\
\hline\\
Optimal detector & $M^m$ \\
MMSE & $m^3+m^2+m M$\\
soft MMSE-SIC \cite{Wang07} & $\mathcal{O}(m^3+m^2+mr^3+mr^2+mM)$\\
GTA \cite{Goldberger11} &  $m^3+ m^2 M$ \\
CHEMP \cite{Narasimhan14} & $rm^2 ~I_{\text{CHEMP}}$ \\
EC (Single L.) & $m^3I_{\text{EC-S}}+m^2 I_{\text{EC-S}}+mMI_{\text{EC-S}}+mI_{\text{EC-S}}$ \\
EC (Double L.) & $m^3DI_{\text{EC-D}}+m^2DI_{\text{EC-D}} + mI_{\text{EC-D}}$ \\\\ \hline
\end{tabular}
\caption{\small Complexity order of different $r\times m$ MIMO detectors. In iterative algorithms, $I_X$ denotes the number of iterations. $D$ is the number of gradient descend steps for the double-loop EC detector. \vspace{0.5cm}} \label{complexity_table}
\end{table}

\noteP{Instead of using small values of $\beta$ to control sudden changes in parameter updates, with the update in \eqref{minvar}, we propose an easy way to artificially control overconfident distributions at early steps of the algorithm, which would restrain the EC algorithm to move far away from the MMSE initial estimate.} We note that using the EC moment matching criterion many other variants of the single loop update methods can be tested and compared with our proposal. However, no significant differences have been appreciated when we measure the system performance in terms of the mutual information in \eqref{eq:mutualinformation} or system bit error rate (BER). In the rest of the paper, regardless of the dimension of the system or constellation order, we implement the EC detector using the single loop approach with $\beta=0.95$, the progressive variance limit in \eqref{minvar} and a maximum number of iterations of $I_{\text{EC-S}}=10$. 

\begin{figure}[t]
\begin{center}
\begin{tabular}{c}
\hspace{-0.5cm}\includegraphics{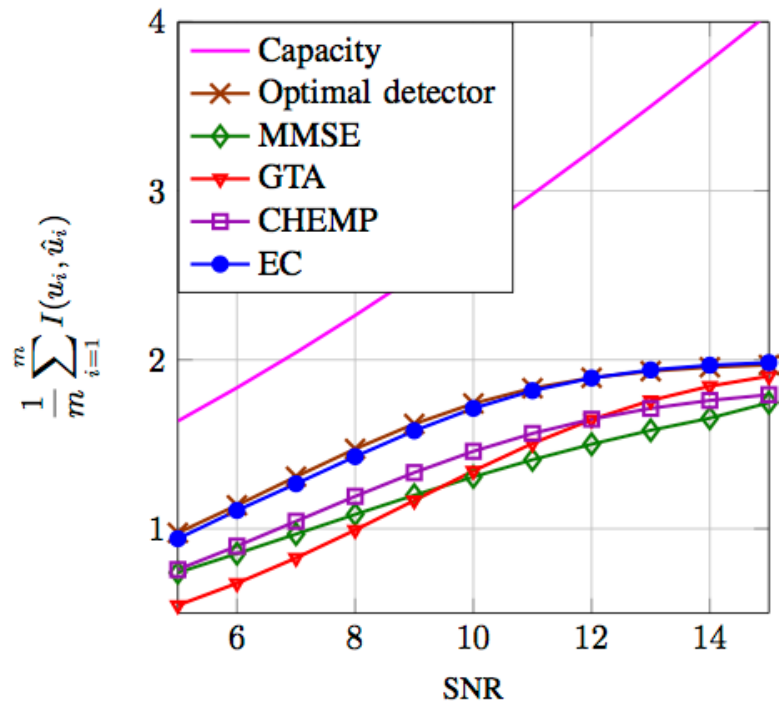}\\ (a) \\ \includegraphics{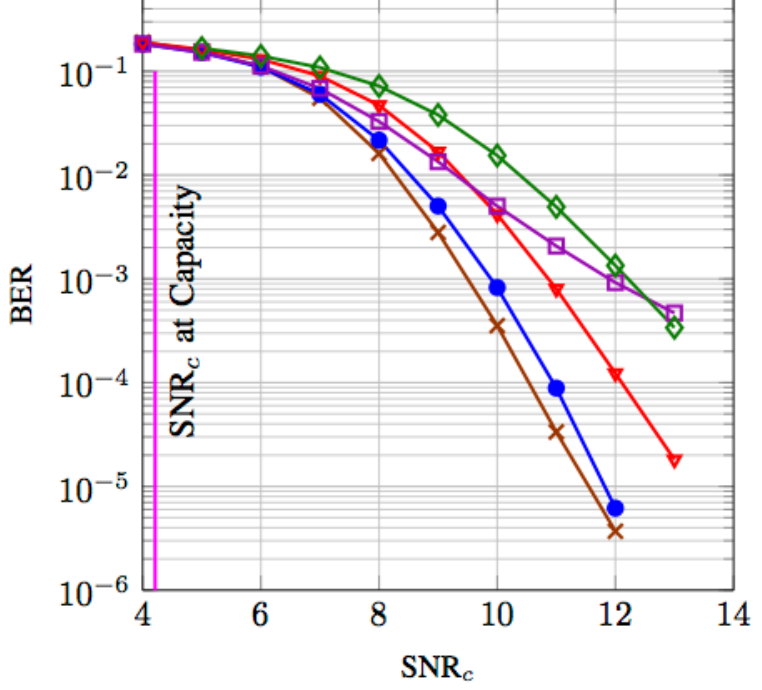}\\
(b)
\end{tabular}
\end{center}
\caption{For an $5 \times 5$ MIMO system with QPSK modulation, in (a) we show the achievable transmission rates. In (b), we include simulated performance when a $(3,6)$-regular LDPC code with block length  $5120$ bits is used.}\label{inf2}
\end{figure}

\begin{figure*}[h!]
\centering
\begin{tabular}{cc}
\multicolumn{2}{c}{\includegraphics{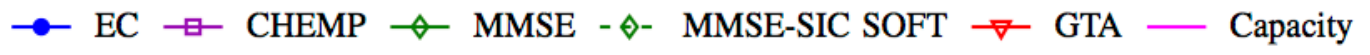}}\\
\hspace{-0.5cm}\includegraphics{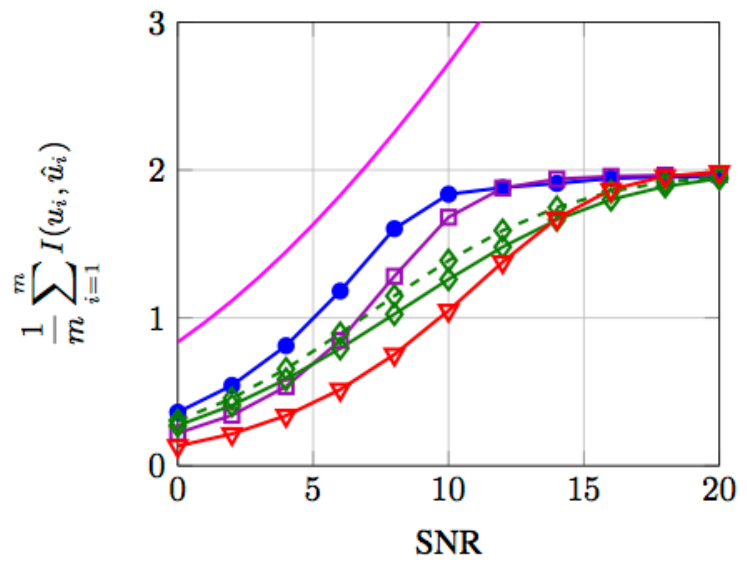} & \includegraphics{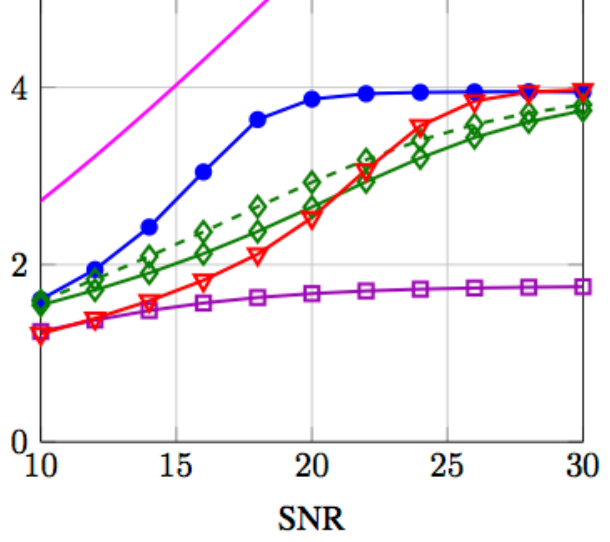} \\  (a) & (b) \\\hspace{0.75cm}\includegraphics{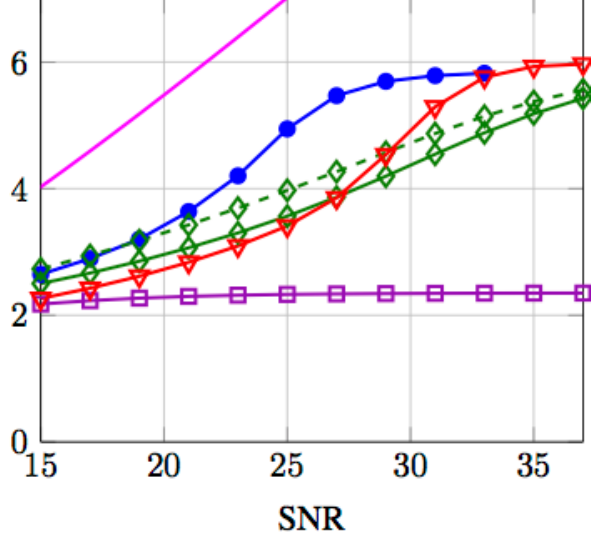} & \includegraphics{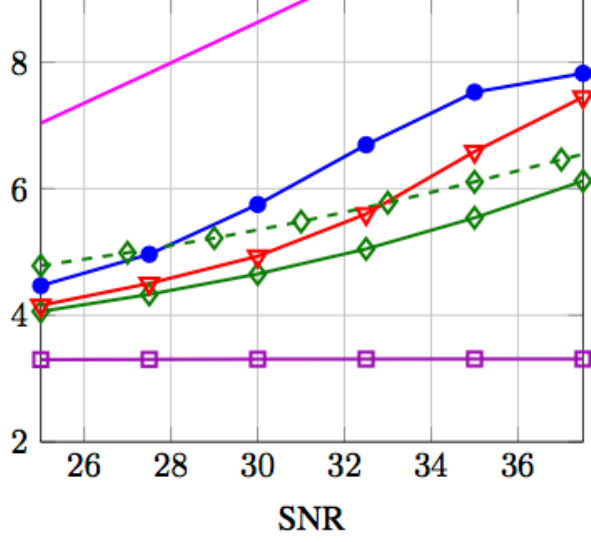}\\
(c) & (d)
\end{tabular}
\caption{Transmission rate computed  for an  $m=r=32$ MIMO system and different constellation orders with QPSK modulation (a), $16$-QAM modulation  (b), $64$-QAM modulation (c) and $256$-QAM modulation (d).}\label{fig:mutualinformation}	   
\end{figure*}

\subsection{Complexity} \label{complexity}

\noteP{In Table \ref{complexity} we summarize the main complexity order of the algorithms presented and those that will be used in our simulation experiments in the next section. In iterative algorithms, $I_X$ denotes the number of iterations. As a rule of thumb, if we run the EC MIMO detector using $I_{\text{EC-S}}=10$ iterations, the incurred complexity is around 10 times larger than the MMSE, GTA and CHEMP complexities.  However, the significant gain in performance that we report in the next section can  justify the increased-complexity of the proposed EC detector.}

\section{Experimental Results} \label{experimental_results}

In the following, we include simulation performance results that demonstrate the accuracy of the EC approximation. In our experiments, we compare  our proposal with the soft output MMSE solution in \cite{Caire-2004,Sanderovich05}, the soft version of the MMSE-SIC in \cite{Wang07}, the GTA algorithm in \cite{Goldberger11}, and the CHEMP method in \cite{Narasimhan14}. \noteP{To avoid cluttering, we do not include in our experiments the GMPID algorithm \cite{Liu16}, since it performs close to CHEMP.} For similar reasons, we do not include the EP method proposed in \cite{Cespedes2014}, since it performs similarly to GTA when used for probabilistic detection \cite{Cespedes14ISIT}.


\subsection{A Low Dimensional MIMO System}

Consider again the  $5\times5$ scenario with QPSK modulation described in Fig. \ref{inf}. Recall that the dimensionality is small enough so we are able to solve the marginalization in \eqref{marg} exactly, which represents the optimal detector. In Fig. \ref{inf2}(a) we include now the results for the EC MIMO detector. Remarkably, it essentially overlaps the optimal  detector performance, achieving a large gain w.r.t. GTA, MMSE and CHEMP. When the number of antennas is small ($5$ in our case), the columns of the channel matrix $\miH$ are typically non-orthogonal and this limits the MMSE performance \cite{Caire-2004,Sanderovich05}. Also, the CHEMP method relies on the matrix $m^{-1}\miH^\top\miH$ being diagonal and for a small $m$, this assumption is unrealistic \cite{Narasimhan14}.

\begin{figure*}
\begin{center}
\begin{tabular}{cc}
\multicolumn{2}{c}{\includegraphics{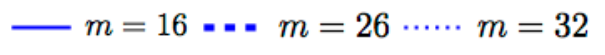}}\\
\hspace{-0.5cm}\includegraphics{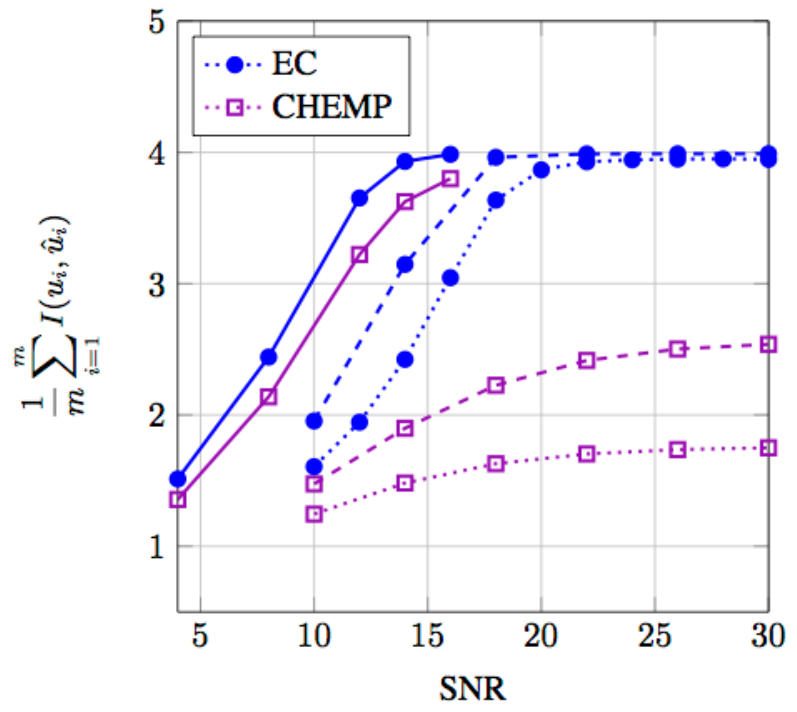} & \includegraphics{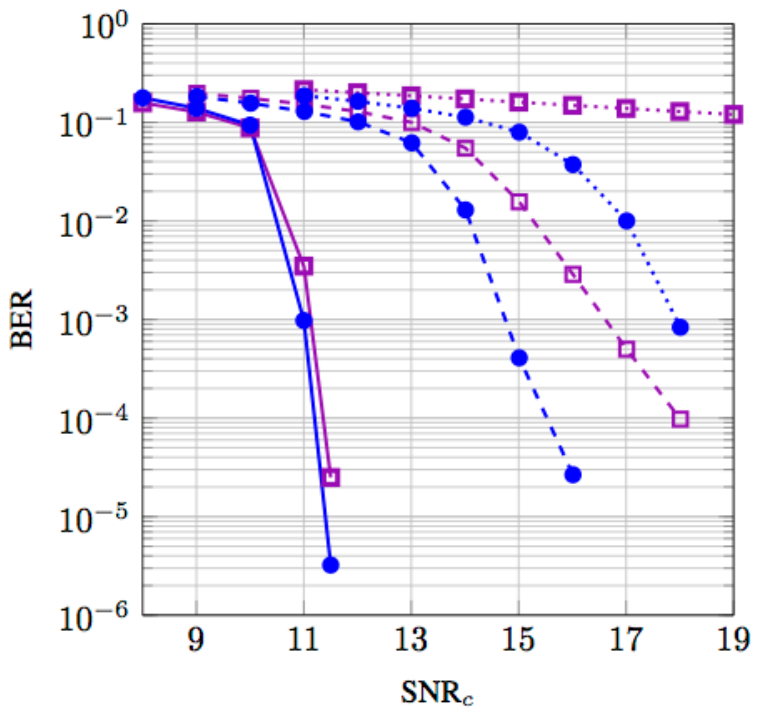}\\
(a) & (b)
\end{tabular}
\end{center}
\caption{For an $32 \times m$ MIMO system with $16$-QAM modulation, in (a) we show the achievable transmission rates for different $m$ values. In (b), we include simulated performance when a $(3,6)$-regular LDPC code with block length  $5120$ bits is used.  }\label{inf3}
\end{figure*}

Results in Fig. \ref{inf2}(a) indicate that the MIMO system performance will highly benefit from the more accurate estimates to the symbol posterior marginals $p(u_i|\y)$ provided by the EC detector. To corroborate this fact, we augment the system model in Fig. \ref{modelo3} by including an LDPC channel encoding stage at the transmitter and an LDPC channel decoder at the receiver. The LDPC channel decoder is fed by soft coded bit probabilities computed using the symbol posterior marginals $p(u_i|\y)$ (or their estimates), according to the bit-modulation mapping. It is well known that the more  accurate the probabilistic detector is, the better performance is obtained after the LDPC decoding stage using BP\cite{Sanderovich05,Olmos2010a,Uchoa14}. In Fig. \ref{inf2}(b), we show for this scenario the simulated BER measured after the LDPC decoding stage (solid lines). A $(3,6)$-regular LDPC code with  block length equal to $5120$ bits has been used. Note that, to simulate the coded performance, the SNR definition in \eqref{snr} is corrected by the coding rate $R$ (the coding rate is $R=0.5$ in the case of $(3,6)$-regular LDPC code). To avoid confusion, we denote this by SNR$_c$, and thus SNR$_c$(dB)=SNR+$10\log_{10}(R)$. 
%
Results have been averaged over 5000 realizations of the channel matrix $\H$.  In terms of coded performance, the gap between optimal detection and EC is only about $0.4$ dB measured at a BER of $10^{-4}$ while the gap to GTA is over $1.5$ dB. In all scenarios observe that, while the soft MMSE-SIC method always improves MMSE, and also GTA al low SNR values, its performance is still far from the EC detector.

\subsection{A $32\times32$ MIMO system}

In a larger scenario, exact marginalization is not viable anymore and we fully rely on approximate methods. In Fig. \ref{fig:mutualinformation}, we represent the obtained achievable rates for a $32\times32$ MIMO system using QPSK modulation (a), $16$-QAM modulation  (b),  $64$-QAM modulation (c), and $256$-QAM modulation (d). While CHEMP and EC are competitive for the QPSK case, CHEMP is no longer a viable option in the $16$-QAM or $64$-QAM cases.  As discussed in \cite{Narasimhan14}, the variance of the interference noise that CHEMP aims to iteratively cancel grows with the constellation order.  For $m=r$ and high order constellations the interference noise becomes excessively large.  Note that the soft MMSE-SIC method always improves MMSE and GTA al low-intermediate SNR values but still its performance is far from the EC detector.



Following \cite{Narasimhan14}, it can be checked that CHEMP becomes effective again as we reduce the number of transmitting antennas, i.e., if $m<r$. In Fig. \ref{inf3} (a), we compare the EC and CHEMP transmission rates for a $16$-QAM modulation with $r=32$ and $m=16$, $26$ and $32$. In (b), we include BER simulation results using the $(3,6)$-regular LDPC code with  block length equal to $5120$ bits. For small $m$ values,  CHEMP  is comparative to the EC solution. However, its performance is severely degraded as $m$ approaches $r$.  CHEMP can be regarded as a Gaussian message-passing distributed implementation of the EC algorithm for those cases where interference is ``locally" tractable. Unlike CHEMP, the EC algorithm performs the update of all parameters at the same time in a centralized manner.  These results show that EC MIMO detector is robust against the increase in the constellation order. In the following we solely consider $m=r$ scenarios with high order constellations and hence we omit CHEMP from the results. 

\begin{figure*}[h]
\begin{center}
\includegraphics{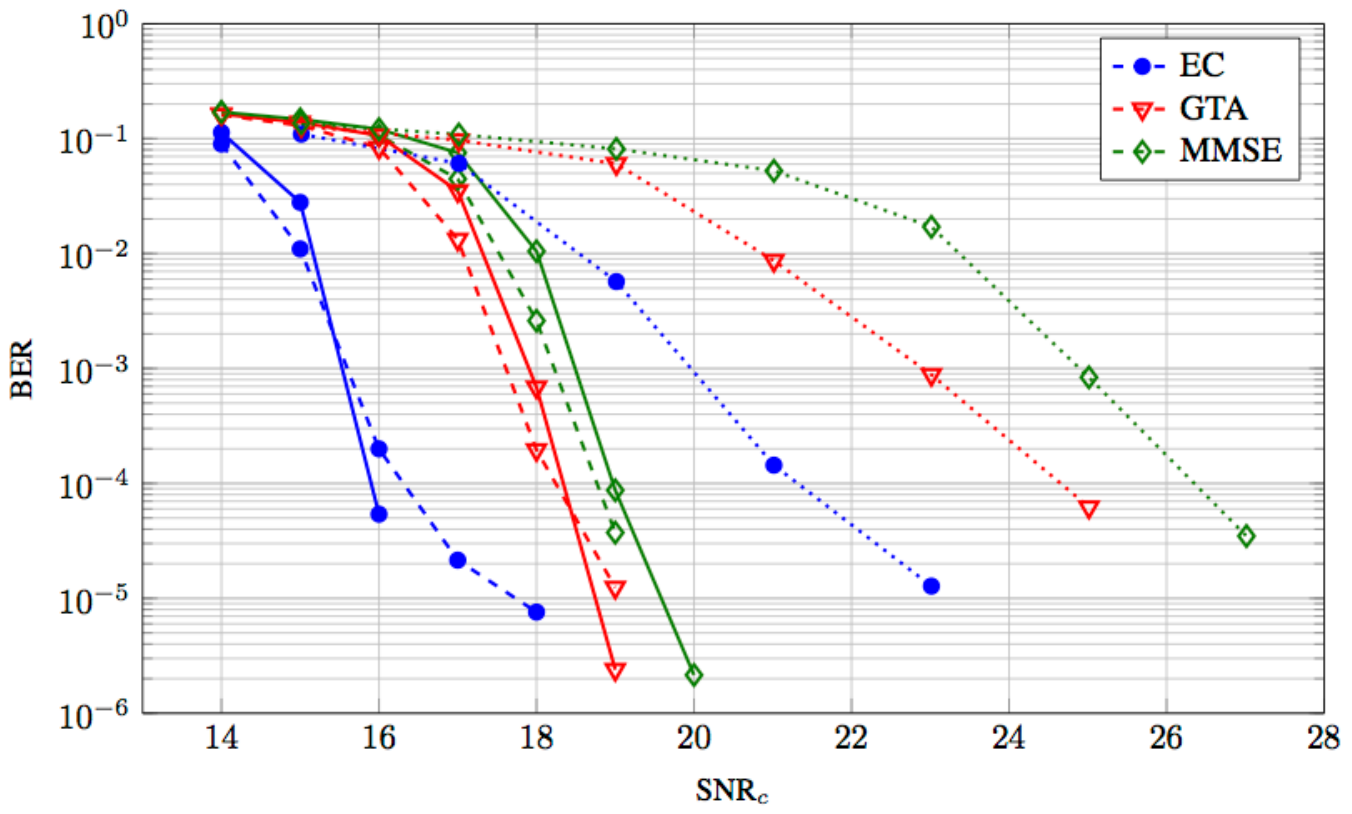}
\caption{System performance of an $32\times32$ $16$-QAM using the irregular rate-$1/2$ LDPC code in [6, Example 3.99] with (dashed lines) block length $30720$ bits, a $(3,6)$-regular LDPC convolutional code  (solid lines) with the same block-length and coding rate $0.479$, \noteP{and a $(3,24)$-regular LDPC convolutional code (dotted lines) with the same block-length and coding rate $0.8698$ \cite{mitchell14}}.}\label{cod_irreg_conv}
\end{center}
\end{figure*}

We complete the study of this scenario by including BER performance results using LDPC constructions that are designed to improve the performance of the $(3,6)$-regular LDPC code used in previous experiments. In Fig. \ref{cod_irreg_conv} with dashed lines we show the performance of the rate-$1/2$ irregular LDPC code in [6, Example 3.99] with block length equal to $30720$ bits. We also include  simulation results (solid lines) for a convolutional LDPC (LDPCC) code constructed by spatially-coupling $48$ independent copies of a $(3,6)$-regular LDPC code, each having block length of $640$ bits,  with low-rate terminations \cite{mplc11}. The resulting coding rate is $0.479$ and the total block length is $30720$ bits.  For the irregular LDPC code, at moderate SNR \noteP{EC} is able to provide a significant gain, which  vanishes at high SNR because of the LDPC error floor. In contrast, because the LDPCC code has large minimum distance, no error floor has been observed in the range of SNR considered and EP achieves a stable gain of $2.5$ dB with respect to GTA.  \noteP{Finally, with dotted lines we include  simulation results for a LDPCC code\footnote{LDPCC codes are generated using protographs \cite{tho03} in order to optimize its minimum distance, as described in \cite{mitchell14}. } with the same block length but constructed by 
spatially-coupling $48$ independent copies of a $(3,24)$-regular LDPC code. The resulting coding rate is $0.869$. }

%

\section{Conclusions}\label{conclusions}

Probabilistic symbol detection is a fundamental problem in high-dimensional MIMO communications since the accuracy of the method employed  to approach the true posterior solution may bring significant performance gains when combined with a modern capacity-approaching channel coding scheme. In this paper we have shown how the EC approximate inference methodology, when applied to the posterior probability distribution of the transmitted symbols, can lead to accurate estimates of the marginal distribution for each transmitted symbol. Further, by computing the average per-antenna mutual information between the transmitted symbols and those distributed according to the EC output, we have shown that the system achievable rate heavily depends on the probabilistic detector accuracy and thus the importance of this stage cannot be diminished by using a more powerful channel code. This is actually corroborated by testing the system performance when we combine the probabilistic output of the symbol detectors with an LDPC channel decoder based on belief propagation. The presented EC probabilistic MIMO detector has cubic complexity with the number of antennas and it is able to greatly improve state-of-the-art methods within only 10 iterations, where a matrix inversion has to be performed per iteration. 

%

\bibliographystyle{ieeetr}
\bibliography{AllBib}

\end{document}